\documentclass[12pt,draftclsnofoot,onecolumn]{IEEEtran}
\ifCLASSINFOpdf
\else
\fi
\usepackage{graphicx}
\usepackage{amsmath}
\usepackage{enumerate}
\usepackage{cite}
\usepackage{amsfonts}
\usepackage{eucal}
\usepackage{amssymb}
\usepackage{epstopdf}
\usepackage[ruled,noline]{algorithm2e}
\usepackage{color}
\usepackage{amsthm}
\newtheorem{thm}{Theorem}

\newtheorem{cor}{Corollary}

\theoremstyle{definition}

\theoremstyle{remark}
\newtheorem{rem}{Remark}

\begin{document}
\title{Hybrid Beamforming Design and Performance with Imperfect Phase Shifters in Multiuser Millimeter Wave Systems}
\author{\IEEEauthorblockN{Wendi Wang, Huarui Yin, Xiaohui Chen, Weidong Wang \\}
\IEEEauthorblockA{Department of Electronic Engineering and Information Science, \\University of Science and Technology of China}
}

\maketitle

\begin{abstract}
 Hybrid beamforming (HBF) includes analog beamforming with phase shifted array in RF domain and digital beamforming in baseband domain. Phase shifted array is usually made up with a large amount of phase shifters. Limited to the manufacturing techniques, phase shifters are unavoidable to have phase-shifting error and gain error. In the paper we study the influence of imperfect phase shifters upon the performance of multiuser mmWave massive MIMO systems\footnote{The analysis of the influence of imperfect phase shifters in point-to-point mmWave massive MIMO systems has been studied in our previous work accepted by 2018 Vehicular Technology Conference {\cite{Previous}}.} with the wireless channel modeled by Rice fading. We derive the upper bound of the achievable sum rate with imperfect phase shifters, which can be very tight when the number of the antennas at the base station is much larger than the number of users. Results show that there is a performance ceiling due to phase-shifting error and gain error. Then we propose a novel channel estimation and hybrid beamforming method to settle this problem with low training overhead. It performs well when the channel paths are few and the channel Rician K-factor is large, which can be easily satisfied in outdoor mmWave communication environment. We further make a tradeoff between the performance and the training overhead so that the algorithm can work well in more kinds of propagation environment.

\end{abstract}

\section{Introduction}

  Future 5G communication systems demand higher data rate, larger bandwidth and higher spectral efficiency. With huge unlicence spectrum, millimeter wave communication from 30-GHz to 300-GHz is considered a promising solution to the lack of spectrum in current communication systems {\cite{work}}. Thanks to the short wavelength in mmWave system, compared with $<6$-GHz radio system, more antennas can be integrated into the same space to get higher directional gain to compensate the path loss {\cite{Model}}. Millimeter wave communication  has been widely used for long-distance point-to-point scenario in satellite and terrestrial applications and is being studied and developed for commercial cellular systems.

  In conventional MIMO systems, one antenna is corresponding to one RF chain (including amplifiers, mixers, ADC/DACs). But in massive mmWave MIMO systems, this may cause unbearable power consumption and hardware cost. Hence hybrid beamforming techniques are introduced into the mmWave massive MIMO systems {\cite{Survey}}.
  Hybrid beamforming (HBF) contains two stages: the analog beamforming (ABF) in the RF domain mainly aimed to reduce the number of RF chains and the digital beamforming (DBF) in the baseband domain functioning as in the conventional MIMO systems. The former is usually implemented with phase shifter network with constant amplitude constraint.

 There are many works coming up with the HBF algorithms for both the point-to-point and multiuser mmWave MIMO systems. In the point-to-point scenario, paper {\cite{OMP}} proposed an OMP based algorithm to jointly design the analog and digital beamforming matrix, which is widely considered as the performance comparison of the subsequent algorithms {\cite{Double,OMP_compare_2,OMP_compare_3}}. In {\cite{SVD}} a two-stage asymptotic optimal hybrid beamforming algorithms based on SVD of the channel matrix was proposed considering the number of RF chains is equal to the number of data streams. The above HBF algorithms are based on the assumption that perfect CSI is available at both the transmitter and the receiver.  With no CSI knowledge, channel estimation should be carried out through the ways of dividing the whole space into beam grids and then applying compressive sensing algorithms {\cite{Grid1,Grid2}} or the off-grid ways applying spatial spectrum method{\cite{MUSIC,SSE}}, gradient descent method {\cite{Gradient}} and so on {\cite{ANM}}.

  In the multiuser scenario, many two-stage HBF algorithms are proposed. To design the analog beamformer, apart from the methods with the requirement of CSI mentioned above {\cite{No_Heq_estimation}}, beam training methods are also widely adopted. Beam training methods are intended to find the best performing candidate vectors by searching among the codebook with different strategies such as maximizing the achievable sum rate, maximizing the data rate of per user {\cite{ZF_important}}, or their combination {\cite{Butler}}. The exhaustive searching performs best but suffers huge computational complexity thus is impractical to be applied. The per-user selection method {\cite{ZF_important}} reduces the computational complexity but the performance is inferior to exhaustive searching. A two-step selection method is proposed {\cite{Butler}} to make a tradeoff between the computational complexity and the performance.
  The DBF is often intended to reduce inter-user interference by, for instance, zero-forcing (ZF) algorithm and block diagonalization (BD) algorithm{\cite{BD_1,BD_2,BD_3}}. It has been proved that applying ZF precoder can achieve linear growth of the sum rate with respect to the number of antennas and users when they go to infinity{\cite{Space_time_multiple_access_Linear_growth_in_the_sum_rate}}, which is approximate to dirty paper coding (DPC). So ZF precoder is widely used as the digital precoder{\cite{ZF_important,ZF_2,ZF_3,No_Heq_estimation}} due to its low complexity and asymptotic optimality.
%

  In RF domain of mmWave massive MIMO systems, the heavy usage of phase shifters (from a few hundred to several thousand) in HBF structure makes the system cost sensitive with the unit price of phase shifters and usually the unit price of mass production determines the component performance. Constrained by cost, volume and production process, it is unavoidable to induce the phase-shifting error and gain error of phase shifters. The rms value of phase-shifting error of the state-of-the-art 360-degree-coverage passive or active phase shifters working at the mmWave frequency is about $0.1^{\circ}$ to $10^{\circ}$ and  the rms value of gain error is about $0.1$ dB to $2$ dB {\cite{PS_28Hz, PS_60Hz,PS_active}}. It is very critical for manufacture to get the relation between the system performance of mmWave massive MIMO systems and the component performance of imperfect phase shifters for mass production. However, to the best of author's knowledge, there is no work analyzing the influence of the phase-shifting error and gain error of phase shifters. Different from the uniformly distributed quantization error of digitally controlled phase shifters considered in {\cite{SVD,ZF_important}}, the phase-shifting error and gain error follow Gaussian distribution, exist in both analog and digital phase shifters and are random and unknown so that they are hard to be compensated.

In our previous work, we have studied the performance loss with imperfect phase shifters in the point-to-point mmWave MIMO systems {\cite{Previous}}. Our analytic procedure therein is applicable for all kinds of hybrid beamforming methods, including, but not limited to the aforementioned two.
  In this paper, we focus on multiuser scenario in the mmWave massive MIMO systems and investigate the performance loss with phase-shifting error and gain error. We only consider the fully-connected hybrid beamforming structure for lack of space and the similar analytic method can be applied to other structures. The main contributions of this paper are listed as follows:
  \begin{itemize}
    \item[1] In the downlink multiuser mmWave massive MIMO systems, we derive the upper bound of the achievable sum rate with the phase-shifting error and gain error. We assume that the base station has all the perfect CSI and each user has its own CSI. We use the ABF design method based on the SVD of channel {\cite{No_Heq_estimation}} as an example and similar analysis can be applied to other ABF design methods. We use ZF precoder as the digital precoder to analyze the achievable sum rate with imperfect phase shifters.
          Both the theoretical derivation and numerical simulation show that when the number of antennas at the BS is far more than the number of users, the upper bound can be very tight. With the phase-shifting error of $5^{\circ}$ and gain error of $1$ dB, the sum rate has nearly $50\%$ degradation when SNR is $25$ dB, which is not negligible. As SNR going larger, the achievable sum rate will be limited to a performance ceiling.

%

    \item[2] To address the aforementioned problem, we proposed a fast and accurate estimation method of the angle-of-departure (AOD) at the downlink transmission and then design the hybrid beamformer. The existing hybrid beamforming methods demand large training overhead to enhance the performance and suffer huge computational complexity thus cannot be carried out in the downlink. Limited to this, many works estimate the CSI in the uplink and design the downlink hybrid beamformer based on the uplink CSI with the assumption of channel reciprocity. But in practice the channel reciprocity is not satisfied. In our proposed downlink  channel estimation algorithm, neither iteration procedure nor searching among the codebook is required thus the training overhead can be cut down greatly, as well as the computational complexity. Then the stage of estimating the equivalent channel is added, based on which the digital precoder can be designed. With the proposed channel estimation and HBF design algorithm, the ceiling of the sum rate in the high SNR regime can be removed and the training overhead can be reduced in the meantime. Our proposed algorithm performs well when the channel paths are few and the channel Rician K-factor is large. We further make a tradeoff between the performance and the training overhead so that the algorithm can work well in more kinds of propagation environment.
  \end{itemize}

  The remainder of the paper is organized as follows. In section \uppercase\expandafter{\romannumeral2}, the system model as well as the channel model is characterized. In section \uppercase\expandafter{\romannumeral3}, we study the loss of the achievable sum rate in multiuser scenario with imperfect phase shifters. Then we proposed an algorithm for the channel estimation and hybrid beamforming design to compensate the performance loss and reduce the training overhead in section \uppercase\expandafter{\romannumeral4}. Simulation results are shown in section \uppercase\expandafter{\romannumeral5} to validate the theorem in the previous section and compare the performance of the existing algorithms with our proposed one. Finally the conclusion is presented in section \uppercase\expandafter{\romannumeral6}. The proof of the theorem is detailed in the appendix.

  {\textbf{Notations:}} In this paper, we use boldface letters to denote matrices(upper case) and vectors(lower case). We use $({\bf{A}})^T$, $({\bf{A}})^*$, $({\bf{A}})^{-1}$, ${\text{tr}}({\bf{A}})$, ${\text{det}}({\bf{A}})$ to denote transpose, Hermitian transpose, inverse, trace and determinant of matrix ${\bf{A}}$, respectively. ${\mathbb{C}}^{M\times N}$ denotes the set of complex-valued matrix with dimension $M\times N$; $||.||_{\text{F}}$ denotes the Frobenius norm of matrix; $E[.]$ denotes the statistical expectation; ${\mathcal{CN}}(\mu,{\bf{R}})$ denotes the complex Gaussian random vector with mean $\mu$ and covariance ${\bf{R}}$; ${\mathcal{N}}(\mu,\sigma^2)$ denotes the real Gaussian random variable with mean $\mu$ and variance $\sigma^2$; ${\emph{\text{Re}}}(a)$ denotes the real part of the complex number $a$.

  \section{System Model}
 \subsection{System Model}
In this section we consider the downlink multiuser mmWave massive MIMO systems. The base station (BS) with $N_{\textrm{BS}}$ antennas communicates to $K$ users simultaneously $N_{\text{BS}} \gg K$. At the BS, the number of RF chains is $N_{\textrm{RF}}$, which is equal to the number of users $K$. At the user side, each user is equipped with $N_{\textrm{UE}}$ antennas and one RF chain.

  At the BS, before transmitted, the symbol vector ${\bf{s}} \in C^{ K \times 1}$ where $E[{\bf{s}}{\bf{s}}^*]={\bf{I}}_{K}$ is precoded by the hybrid precoder ${\bf{F}} = {\bf{F}}_{\textrm{RF}} {\bf{F}}_{\textrm{BB}} $ where ${\bf{F}}_{\textrm{RF}} \in {\mathbb{C}}^{N_{\textrm{BS}} \times K}$ is the analog precoder and ${\bf{F}}_{\textrm{BB}} = [{\bf{f}}_1^{\textrm{BB}},{\bf{f}}_2^{\textrm{BB}},\cdots,{\bf{f}}_K^{\textrm{BB}}]\in {\mathbb{C}}^{K \times K}$ denotes the digital precoder in the baseband domain. The analog precoding matrix is usually implemented with phase shifters so the elements are limited to the constant amplitude constraint $|f^{\text{RF}}_{i,j}|^2 = N_{\text{BS}}^{-1},i=1,2,\cdots,N_{\text{BS}},j=1,2,\cdots,K$. Then the transmitted signal vector is denoted as ${\bf{x}} = {\bf{F}}_{\textrm{RF}} {\bf{F}}_{\textrm{BB}} {\bf{s}}$.

 The received signal vector at the $k$-th user is given as below
  \begin{equation}\label{raw receive signal}
  {\bf{r}}_k = {\sqrt{\rho}}{\bf{H}}_k {\bf{F}}_{\textrm{RF}} \sum_{u=1}^{K} \eta {\bf{f}}_u^{\textrm{BB}} s_u + {\bf{z}}_k
  \end{equation}
  where $\rho$ indicates the average transmitted power for each user, ${\bf{H}}_k \in {\mathbb{C}}^{N_{\text{UE}} \times N_{\text{BS}}}$ denotes the channel state information from the BS to the $k$-th user, which will be modeled later and ${\bf{z}}_k \sim {\mathcal{CN}}({\bf{0}}_{N_{\text{UE}} \times 1}, \sigma^2_z {\bf{I}}_{N_{\text{UE}}})$ is the noise vector with independent and identically distributed (i.i.d.) elements. It is assumed that the BS has the perfect CSI of all the users and each user has its own CSI. The factor $\eta = \sqrt{\frac{1}{{\textrm{tr}} ( {\bf{F}}_{\textrm{BB}} {\bf{F}}_{\textrm{BB}}^*)}}$ is the normalization factor of transmission power. At the user side, only analog combiner is adopted for there is only one RF chain, which is denoted as ${\bf{w}}_k \in {\mathbb{C}}^{N_{\text{UE}} \times 1}$ and $|w_{ki}|^2 = N_{\text{UE}}^{-1},i=1,2,\cdots,N_{\text{UE}}$. After analog combining, the received signal of the $k$-th user can be written as

\begin{small}
  \begin{equation}\label{receive siganl after analog combining}
  y_k = \underbrace{\sqrt{\rho} {\bf{w}}_k^* {\bf{H}}_k {\bf{F}}_{\textrm{RF}} \eta {\bf{f}}_k^{\textrm{BB}} s_k }_{\text{desired signal}}  + \underbrace{\sqrt{\rho} {\bf{w}}_k^* {\bf{H}}_k {\bf{F}}_{\textrm{RF}} \sum_{u \neq k} \eta {\bf{f}}_u^{\textrm{BB}} s_u }_{\text{interference signal}} + \underbrace{{\bf{w}}_k^* {\bf{z}}_k }_{\text{noise item}}.
  \end{equation}
\end{small}

In this paper, the fully-connected hybrid beamforming is considered where one RF chain is connected with all the antennas through the phase shifter network. The phase shifters can be analog with continuous phase or digital with $B$-bit resolution discrete phase. With the phase shifter network, the analog precoding matrix can be expressed as

\begin{equation}\label{Frf}
  {\bf{F}}_{\textrm{RF}} = \frac{1}{\sqrt{N_{\text{BS}}}} \left(
                             \begin{array}{cccc}
                               e^{j \theta_{11}} & e^{j \theta_{12}} & \cdots & e^{j \theta_{1N_{\textrm{RF}}}} \\
                               e^{j \theta_{21}} & e^{j \theta_{22}} & \cdots & e^{j \theta_{2N_{\textrm{RF}}}} \\
                               \vdots & \vdots & \ddots & \vdots \\
                               e^{j \theta_{N_{\textrm{T}}1}} & e^{j \theta_{N_{\textrm{T}}2}} & \cdots & e^{j \theta_{N_{\textrm{T}} N_{\textrm{RF}}}} \\
                             \end{array}
                           \right)
\end{equation}
  where $\forall \theta _{n_{\text{T}} n_{\textrm{RF}}} \in \Theta, n_{\text{T}} \in \left\{ 1,\cdots,N_{\textrm{T}} \right\}, n_{\textrm{RF}} \in \left\{ 1,\cdots,N_{\textrm{RF}} \right\}$ and $\Theta = [0,2\pi / 2^B,\cdots,(2^B-1)2\pi/2^B]$ for digital $B$-bit resolution phase shifters, $\Theta = [0,2\pi] $ for analog phase shifters. The analog combiner is the same.

  When there exist phase-shifting error and gain error, the elements of analog beamforming matrix will become  $\frac{1}{\sqrt{N_{\text{BS}}}}\alpha _{n_{\text{T}} n_{\text{RF}}} e^ {j(\theta _{n_{\text{T}} n_{\textrm{RF}}} + \delta  _{n_{\text{T}} n_{\text{RF}}})}$ with the gain error $\alpha _{n_{\text{T}} n_{\text{RF}}} \sim {\mathcal{N}}(1,\sigma _\alpha^2)$ and the phase-shifting error $\delta  _{n_{\text{T}} n_{\text{RF}}} \sim {\mathcal{N}}(0,\sigma _\delta^2)$. And the analog precoding matrix, the analog combining vector, and the hybrid precoding matrix are denoted by ${\bf{F}} _{\text{RF,E}}, {\bf{w}} _{\text{E}},{\bf{F}} _{\text{E}}$, separately. Since the phase-shifting error and gain error are unknown to the BS, they can't be compensated in the baseband domain and the practical digital beamforming matrix ${\bf{F}}_{\text{BB}}$ has no relation with the phase-shifting error and gain error.

 \subsection{Channel Model}
In the multiuser scenario, the user equipment may have very few antennas thus only the LOS path is dominant and the scattering paths can be negligible in the long-distance propagation or non-negligible but very weak relatively in the short-distance propagation referring to {\cite{ZF_important}}. Therefore, we take the Rician fading channel with a large Rician K-factor into account. The channel of the $k$-th user can be expressed as

\begin{equation}\label{Channel_Rice}
  {\bf{H}}_k =  \sqrt{\frac{v_k}{v_k + 1}} {\bf{H}}_{\text{L},k} +  \sqrt{\frac{1}{v_k + 1}} {\bf{H}}_{\text{S},k}
\end{equation}
where $v_k$ is the Rician K-factor of the $k$-th user and we set $v_k = v$, $\forall k$ to simplify the analysis. ${\bf{H}}_{\text{L},k}$ is the channel matrix of LOS path and ${\bf{H}}_{\text{S},k}$ is the NLOS component. Adopting the uniform linear array (ULA) structure, we have
\begin{equation}\label{Channel_los_nlos}
\begin{aligned}
 {\bf{H}}_{\text{L},k} & =  {\sqrt{N_{\text{BS}} N_{\text{UE}}}} {\bf{a}}_{{\text{UE}}} (\phi^{{\text{UE,L}}}_{k}) {\bf{a}}_{{\text{BS}}}^* (\phi^{{\text{BS,L}}}_{k}) \\
 {\bf{H}}_{\text{S},k} & = {\sqrt{\frac{N_{\text{BS}} N_{\text{UE}}}{L}}} \sum_{l=1}^{L} \gamma_{k,l} {\bf{a}}_{{\text{UE}}} (\phi^{{\text{UE,S}}}_{k,l}) {\bf{a}}_{{\text{BS}}}^* (\phi^{{\text{BS,S}}}_{k,l})
\end{aligned}
\end{equation}
where $\gamma_{k,l} \sim {\mathcal{CN}}(0,1)$ is the path coefficient of the $l$-th scattering path corresponding to the $k$-th user. The angles $\phi^{{\text{UE,L}}}_{k}$, $\phi^{{\text{BS,L}}}_{k}$ represent the downlink angle-of-departure (AOD) and angle-of-arrival (AOA) of the LOS path and $\phi^{{\text{UE,S}}}_{k,l}$,$\phi^{{\text{BS,S}}}_{k,l}$ stand for the downlink AOD and AOA of the $l$-th scattering path. $ {\bf{a}}_{\text{BS}} (*)$ and $ {\bf{a}}_{\text{UE}} (*)$ represent the antenna array response vectors at the BS and user side. The $N$-dimensional antenna array response vector at angle $\phi$, for ULA structure is expressed as

\begin{equation}\label{antenna array response vector}
  {\bf{a}} (\phi) = \frac{1}{\sqrt{N}}(1, e^{\frac{j2\pi d}{\lambda} \text{cos} (\phi)}, \cdots, e^{\frac{j2\pi d}{\lambda} (N-1) \text{cos} (\phi)})^T
\end{equation}
 where $\lambda$ is the wavelength and $d$ is the antenna spacing. In this paper, it is assumed that the transmitter and receiver are equipped with linear arrays with $d = \lambda / 2$.

\section{Performance loss in Multiuser scenario}
In this section we consider the performance loss due to the phase-shifting error and gain error in the downlink multiuser mmWave MIMO system. We use the asymptotically optimal hybrid beamformer for the multiuser scenario. Firstly the SVD factorizes each user's channel matrix as ${\bf{H}}_k = {\bf{U}}_k{\bf{\Sigma}}_k{\bf{V}}_k^*$ and the diagonal elements of ${\bf{\Sigma}}_k$ is in a descending order. The analog precoder at the BS is ${\bf{f}}_k^{\textrm{RF}} = e^{j\mathcal{Q}(\angle {\bf{V}}_{k,[:,1]})}$ and the analog combiner at each user is ${\bf{w}}_k = e^{j\mathcal{Q}(\angle {\bf{U}}_{k,[:,1]})}$ where ${\mathcal{Q}}(\cdot)$ is the quantization of the angle in the parentheses according to the resolution of the phase shifters. When the number of users is large, more than one users share a very similar AOD with a high probability to some extent and these users may cause large interference to each other. In this case, the user scheduling is indispensable and was ignored in most of the existing research. The work in {\cite{ADMA}} proposed a greedy angle division multiple access (ADMA) user scheduling algorithm to divide the users into different scheduling groups. In our work we use a simple way that the users within a AOD range of $2 \pi /N_{\text{BS}}$ rad will be silent in the current transmission cycle except only one user. And the analog beamforming vectors corresponding to the silent users will be set as zero vectors. We use ${\mathnormal{I}}$ to denote the index set of the non-silent users and its cardinality is $K_{\mathnormal{I}}$ ($K_{\mathnormal{I}} \leq K$). The analog beamforming matrix at the BS can be written as ${\bf{F}}_{\text{RF}} = [{\bf{f}}_{I_1}^{\textrm{RF}}, {\bf{f}}_{I_2}^{\textrm{RF}} ,\cdots,{\bf{f}}_{I_{K_I}}^{\textrm{RF}}]^T$ where $I_{k_I}$ stands for the $k_I$-th element in the set $I$, $k_I = 1,2,\cdots,K_I$.

As to the digital precoder at the BS, the dirty paper coding (DPC) is optimal but too difficult to implement. The low complexity linear precoding algorithms such as zero-forcing (ZF) are preferred whose performance converge to the optimal DPC in the massive MIMO systems. The digital ZF precoder at the BS is given by

  \begin{equation}\label{ZF precoder}
    {\bf{F}}_{\textrm{BB}} = {\bf{H}}_{\textrm{eq}}^*({\bf{H}}_{\textrm{eq}} {\bf{H}}_{\textrm{eq}}^*)^{-1}
  \end{equation}
where the equivalent channel matrix can be expressed as {\cite{ZF_important}}
  \begin{equation}\label{Heq}
    {\bf{H}}_{\textrm{eq}} =
    \begin{bmatrix}
    {\bf{w}}_{I_1}^*  {\bf{H}}_{I_1}  {\bf{F}}_{\textrm{RF}} \\ \vdots \\{\bf{w}}_{I_{K_I}}^*  {\bf{H}}_{I_{K_I}}  {\bf{F}}_{\textrm{RF}}
    \end{bmatrix}
    = [ {\bf{h}}_{{\textrm{eq}},1} \quad {\bf{h}}_{{\textrm{eq}},2} \cdots {\bf{h}}_{{\textrm{eq}},K_I} ]^T
  \end{equation}
  where ${\bf{h}}_{{\textrm{eq}},k_I} = {\bf{w}}_{I_{k_I}}^*  {\bf{H}}_{I_{k_I}} {\bf{F}}_{\textrm{RF}}$.

  With the ZF precoder, the multiuser interference can be eliminated and the SINR of the $k_I$-th non-silent user can be expressed as

  \begin{equation}\label{SINR NO ERROR}
    {\textrm{SINR}}_{k_I} = \frac{\eta^2 \rho}{\sigma_{\textrm{UE}}^2}
  \end{equation}

  Then the achievable sum rate is
  \begin{equation}\label{rate no error}
    R_{\textrm{HB}} = K_I \cdot {\textrm{log}}_2 \left\{ 1 + \frac{\eta^2 \rho}{\sigma_{\textrm{UE}}^2}   \right\}
  \end{equation}

  In the sequel, the phase-shifting error and gain error of phase shifters is taken into account and the equivalent channel under these error ${{\bf{H}}}_{\textrm{eq,E}}$ can be expressed as
  \begin{equation}\label{Heq error}
  \begin{aligned}
  \begin{split}
      {{\bf{H}}}_{\textrm{eq,E}}
      & =
    \begin{bmatrix}
    {\bf{w}}_{{\text{E}},I_1}^*  {\bf{H}}_{I_1}  {\bf{F}}_{\textrm{RF,E}} \\ \vdots \\{\bf{w}}_{{\text{E}},I_{K_I}}^*  {\bf{H}}_{I_{K_I}}  {\bf{F}}_{\textrm{RF,E}}
    \end{bmatrix}
    = [ {\bf{h}}_{{\textrm{eq}},1}^{\text{E}} \quad {\bf{h}}_{{\textrm{eq}},2}^{\text{E}} \cdots {\bf{h}}_{{\textrm{eq}},{K_I}}^{\text{E}} ]^T \\
    &={\bf{H}}_{\textrm{eq}} + \Delta {\bf{H}}_{\textrm{eq}}\\
  \end{split}
  \end{aligned}
  \end{equation}
 where $\Delta {\bf{H}}_{\textrm{eq}} = [ \Delta {\bf{h}}_{{\textrm{eq}},1} \quad \Delta {\bf{h}}_{{\textrm{eq}},2} \cdots \Delta {\bf{h}}_{{\textrm{eq}},{K_I}} ]^T $.

  We use ${\bf{F}}_{\textrm{BB}}$ and ${{{\bf{F}}}}_{\textrm{BB,E}}$ to define the digital ZF precoder based on ${\bf{H}}_{\textrm{eq}}$ and ${{\bf{H}}}_{\textrm{eq,E}}$ respectively. Note that ${{\bf{H}}}_{\textrm{eq,E}}$ is the equivalent channel in practice but unknown due to the phase-shifting error and the gain error of phase shifters. What we know is ${\bf{H}}_{\textrm{eq}}$ consisting of perfect phase shifters and we can only design the digital precoder according to ${\bf{H}}_{\textrm{eq}}$, instead of ${{\bf{H}}}_{\textrm{eq,E}}$. Thus the received signal at the $k_I$-th user is given by
 \begin{equation}\label{receive siganl after analog combining with error}
  \begin{aligned}
  \begin{split}
  y_{k_I} &= \sqrt{\rho}{{\bf{h}}}_{{\textrm{eq}},{k_I}}^{\text{E}} \eta {\bf{f}}_{k_I}^{\textrm{BB}} s_{k_I} + \sqrt{\rho} {{\bf{h}}}_{{\textrm{eq}},{k_I}}^{\text{E}} \sum_{u \neq {k_I}} \eta {\bf{f}}_u^{\textrm{BB}} s_u + {\bf{w}}_{k_I}^* {\bf{z}}_{k_I} \\
   &= \sqrt{\rho}({\bf{h}}_{{\textrm{eq}},{k_I}} + \Delta {\bf{h}}_{{\textrm{eq}},{k_I}}) \eta {\bf{f}}_{k_I}^{\textrm{BB}} s_{k_I} \\
    & \quad + \sqrt{\rho} ({\bf{h}}_{{\textrm{eq}},{k_I}} + \Delta {\bf{h}}_{{\textrm{eq}},{k_I}}) \sum_{u \neq {k_I}} \eta {\bf{f}}_u^{\textrm{BB}} s_u + {\bf{w}}_{k_I}^* {\bf{z}}_{k_I}  \\
    & = \sqrt{\rho} \eta s_{k_I} + \sqrt{\rho} \eta  \Delta {\bf{h}}_{{\textrm{eq}},{k_I}} {\bf{F}}_{\textrm{BB}} {\bf{s}}_I + {\bf{w}}_{k_I}^* {\bf{z}}_{k_I}  \\
    \end{split}
    \end{aligned}
  \end{equation}

  From ({\ref{receive siganl after analog combining with error}}), we can see the multiuser interference can't be eliminated due to the mismatch of digital precoder ${\bf{F}}_{\text{BB}}$ and the practical equivalent channel ${{\bf{H}}}_{\textrm{eq,E}}$. In this case, the SINR of the $k_I$-th user can be written as
\begin{equation}\label{SINR ERROR}
    \begin{aligned}
  \begin{split}
    {\tilde{\textrm{SINR}}}_{k_I} & = \frac{\eta ^2 \rho}{\eta ^2 \rho x_{k_I} + \sigma_{\textrm{UE}}^2} \\
        \end{split}
    \end{aligned}
  \end{equation}
 where we define $x_{k_I} = \Delta {\bf{h}}_{{\textrm{eq}},{k_I}} {\bf{F}}_{\textrm{BB}}{\bf{F}}_{\textrm{BB}}^* \Delta {\bf{h}}_{{\textrm{eq}},{k_I}} ^* $ for the sake of simplifying the notation.

  Then the achievable sum rate under phase-shifting error and gain error of phase shifters is
\begin{equation}\label{rate_error}
    \begin{aligned}
  \begin{split}
    & R_{\textrm{HB,error}} = \\
    & K_I \cdot \int {\textrm{log}}_2 \left\{ 1 + \frac{\eta ^2 \rho}{\eta ^2 \rho x_{k_I} + \sigma_{\textrm{UE}}^2}   \right\} f(x_{k_I}) dx_{k_I} \\
        \end{split}
    \end{aligned}
  \end{equation}
  where the normalization factor $\eta = \sqrt{\frac{1}{{\text{tr}}({\bf{F}}_{\textrm{BB}}{\bf{F}}_{\textrm{BB}}^*)}}$.

  However, it is hard to derive the explicit result of ({\ref{rate_error}}). In {\cite{ZF_important}}, an approximate result of the achievable sum rate ({\ref{rate_error}}) is given by

  \begin{equation}\label{rate error_mean}
    \begin{aligned}
  \begin{split}
    & R_{\textrm{HB,error}} =
     {K_I} \cdot {\textrm{log}}_2 \left\{ 1 + \frac{\eta ^2 \rho}{\eta ^2 \rho E_{\Delta {\bf{h}}_{{\textrm{eq}},{k_I}} }[x_{k_I} ] + \sigma_{\textrm{UE}}^2}   \right\} \\
        \end{split}
    \end{aligned}
  \end{equation}
   By calculating the second derivative of the function ${\text{log}}_2 \left(1+\frac{a}{ax+b} \right)$, we find that when $a,x,b>0$, the function is convex and the formula ({\ref{rate error_mean}}) is the lower bound of ({\ref{rate_error}}) exactly. In massive MIMO systems, formula ({\ref{rate error_mean}}) can be very approximate to ({\ref{rate_error}}) as shown in Fig. {\ref{aMultiuser_13_14}} in the section \uppercase\expandafter{\romannumeral5} thus formula ({\ref{rate error_mean}}) can describe the effect of phase-shifting error and gain error upon the achievable sum rate well. Then we calculate the theoretical value of the expectation of $x_{k_I}$ in theorem 1.

\begin{thm}
In multiuser scenario, the lower bound of $ E_{\Delta {\bf{h}}_{{\textrm{eq}},{k_I}} }[\Delta {\bf{h}}_{{\textrm{eq}},{k_I}} {\bf{F}}_{\textrm{BB}}{\bf{F}}_{\textrm{BB}}^* \Delta {\bf{h}}_{{\textrm{eq}},{k_I}} ^* ] $ can be expressed as
     \begin{equation}\label{lower bound final}
  \begin{aligned}
  \begin{split}
 E_{\Delta {\bf{h}}_{{\textrm{eq}},k_I} }[x_{k_I}] & \geqslant (\sigma_{\delta}^2)^2 + \frac{K_I (\sigma_{\delta}^2 + \sigma_{\alpha}^2)^2}{N_{\text{BS}}N_{\text{UE}}} \\
& \quad +(\sigma_{\delta}^2+\sigma_{\alpha}^2) e^{-\sigma_{\delta}^2}(\frac{K_I}{N_{\text{BS}}}+\frac{1}{N_{\text{UE}}}) \\
  \end{split}
  \end{aligned}
  \end{equation}
 When $N_{\text{BS}} \to \infty$ or $N_{\text{BS}} \gg K_I$, the bound can be very tight.
\end{thm}

 {\flushleft{\emph{Proof:} Refer to Appendix A.}}

 Thus we can derive the lower bound of the loss of achievable sum rate based on the expression

     \begin{equation}\label{Delta_RRR}
  \begin{aligned}
  \begin{split}
\Delta R \approx {K_I} \cdot {\text{log}}_2\left\{ 1+ \frac{\eta^2 \rho E_{\Delta {\bf{h}}_{{\textrm{eq}},{k_I}} }[x_{k_I}] }{\sigma^2_{\text{UE}}}\right\}
  \end{split}
  \end{aligned}
  \end{equation}

In the low SNR regime where $\frac{\sigma^2_{\text{UE}}}{\eta^2 \rho} \gg E_{\Delta {\bf{h}}_{\text{eq},k_I}}[x_{k_I}]$, the loss of the achievable sum rate due to the phase-shifting error and gain error can be neglected. In the high SNR regime, we further calculate the performance ceiling of the achievable sum rate which is presented in the following corollary.

\begin{cor}
In the high SNR regime, the performance ceiling of the achievable sum rate with imperfect phase shifters is
\begin{equation}{\label{Ceiling}}
R^{\text{ceiling}}_{\text{HB,error}} \approx \sum p(K_I) \cdot K_I {\text{log}}_2 \left\{ 1+ \frac{1}{{\text{LB}}} \right\}
\end{equation}
where `LB' represents the lower bound of $E_{\Delta {\bf{h}}_{\text{eq},k_I}}[x_{k_I}]$ given in ({\ref{lower bound final}}). Formula ({\ref{Ceiling}}) shows that the achievable sum rate is limited to a ceiling only determined by the variance of phase-shifting error and gain error of phase shifters.
\end{cor}

 {\flushleft{\emph{Proof:} As the SNR going infinity, $\frac{\sigma^2_{\text{UE}}}{\eta^2 \rho} \to 0$ thus we have $ \frac{\eta ^2 \rho}{\eta ^2 \rho E_{\Delta {\bf{h}}_{{\textrm{eq}},{k_I}} }[x_{k_I} ] + \sigma_{\textrm{UE}}^2} =  \frac{1}{ E_{\Delta {\bf{h}}_{{\textrm{eq}},{k_I}} }[x_{k_I} ] + \frac{\sigma_{\textrm{UE}}^2}{\eta ^2 \rho}} \to \frac{1}{E_{\Delta {\bf{h}}_{\text{eq},k_I}}[x_{k_I}]}$. $p(K_I)$ is the probability of the case that the number of the non-silent users is $K_I$. It is hard to calculate the explicit expression of $p(K_I)$ but we can obtain the value of $p(K_I)$ by Monte-Carlo simulation. As an example, when $N_{\text{BS}} = 128$ and $K=10$, we have $p(10) = 0.48$, $p(9) = 0.38$, $p(8) = 0.12$, $p(7) = 0.02$. And for simplicity we can just use

\begin{small}
\begin{equation}\label{Upperbound_128_10}
\begin{split}
R^{\text{ceiling}}_{\text{HB,error}} &\approx p(10) \cdot 10{\text{log}}_2 \left\{ 1+ \frac{1}{{\text{LB}}} \right\} + p(9) \cdot 9{\text{log}}_2 \left\{ 1+ \frac{1}{{\text{LB}}} \right\}  \\
& \quad + (1-p(10)-p(9))\cdot 8 {\text{log}}_2 \left\{ 1+ \frac{1}{{\text{LB}}} \right\} \\
\end{split}
\end{equation}
\end{small}

as an approximation of the performance ceiling.   $\hfill\blacksquare$}}

\begin{rem}
In this paper, we focus on the ZF digital precoder at the BS and analyze the performance loss with imperfect phase shifters. Other digital precoders like regularized ZF (RZF) which can further improve the performance in the low SNR regime can be analyzed by the similar procedure. The RZF digital precoder can be expressed as {\cite{RZF}}
 \begin{equation}
   {\bf{F}}_{\text{BB,RZF}} = {\bf{H}}_{\text{eq}}^* ({\bf{H}}_{\text{eq}} {\bf{H}}_{\text{eq}}^* + K_I\frac{ \sigma^2_z}{\rho} {\bf{I}}_{K_I})^{-1}
 \end{equation}
where the transmission power normalization factor $\eta_{\text{RZF}} = \frac{1}{\sqrt{{\bf{F}}_{\text{BB,RZF}}{\bf{F}}_{\text{BB,RZF}}^*}}$. It is easy to prove that $\eta^2 E_{\Delta {\bf{h}}_{\text{eq},k_I}}[x_{k_I}]$ is equal applying RZF and ZF digital precoder with the same analytic procedure. Thus the loss of the achievable sum rate caused by phase-shifting error and gain error is nearly the same with RZF or ZF precoder. While $\eta_{\text{RZF}} > \eta_{\text{ZF}}$ thus the achievable sum rate will be improved with RZF precoder especially in the low SNR regime.
\end{rem}

\begin{rem}
When the number of users communicating to the BS simultaneously is large, the user scheduling scheme will not be optimal of course. Some hybrid beamforming algorithms will achieve higher sum rate such as the exhaustive search among large enough amount of candidate analog beamforming vectors with the strategy of maximizing the sum rate. In these cases, the analytic procedure is also applicable, as well as the derived lower bound. In fact as long as the digital precoder is the linear precoder such as maximum ratio transmission (MRT), ZF, RZF, the similar analytic procedure can be used to different analog beamforming algorithms. For other digital precoder design such as block diagonalization (BD), we will leave it to the future research.
\end{rem}
%
%

The reason of the performance loss is mainly two-fold. One is that the main lobe width of the directional beam will be larger and the side lobe level will be higher due to the distorted analog beamforming vectors with the imperfect phase shifters, which indicates that the beam power will dissipate more and the directional gain will be lower. The other is the ZF digital precoder cannot eliminate the inter-user interference since they lack the perfect knowledge of the practical equivalent channel which is the combination of the analog beamforming matrices at the BS and users and the propagation channel. The latter is easy to deal with by estimating the practical equivalent channel. However the former is hard to calibrate because it is impractical to get the exact knowledge of each imperfect phase shifter for the huge complexity. Fortunately, the former one causes far less harm than the latter one and we just use the imperfect phase shifted array as the analog beamforming vectors. In the next section, we will propose a channel estimation and hybrid beamforming method to resist the phase-shifting error and gain error.

\section{Channel estimation and hybrid beamforming design}
The massive MIMO systems face the challenge of the huge training overhead of channel estimation in the downlink as well as the heavy feedback overhead. Furthermore, the user equipment will bear the heavy burden of high computational complexity to estimate the downlink channel, which is not a wise choice for power saving. To deal with it, most of the existing approaches prefer TDD mode with the assumption of channel reciprocity thus the channel can be estimated in the uplink and the downlink precoding matrix can be designed according to the uplink CSI. However, it is worth mentioning that although the propagation channel is reciprocal, the transceiver RF chains are not and we need calibration methods to deal with the non-reciprocity, which incurs extra overhead or hardware cost. In this section, we find a way to estimate the channel in the downlink with low training overhead and limited feedback. The user equipment can also enjoy the low computational complexity for saving power. No reciprocity calibration is required and both TDD and FDD modes can be applied. What's more, we can design the hybrid beamformer to remove the performance ceiling due to the imperfect phase shifter.

Considering hybrid mmWave MIMO systems under Rice channel model, the main idea of our proposed algorithm is that the analog precoder is intended to steer the beam at the LOS path direction and the power of other scattering paths can be collected partially by digital precoder when the LOS path is far stronger than the scattering paths. Thus to design the analog precoder at the BS, the knowledge of the downlink AOD corresponding to the LOS path is enough, instead of the whole knowledge of channel. Once we have the analog precoder, the BS can send training signals to the users to get the knowledge of the equivalent channel, which has much fewer elements and lower dimension compared with the propagation channel and contains the contribution of scattering paths. Then the digital precoder can be designed based on the equivalent channel. The whole process consists of four stages as shown in Algorithm 1.

At the first stage, it is intended to estimate the downlink AOD of the LOS path at the user side. The DFT interpolation method is applied here to estimate the AOD by directly interpolating the complex-valued DFT coefficients. The BS sends the pilot matrix ${\bf{S}} = \sqrt{\rho} {\bf{I}}_K$ using $K$ time slots and the analog precoding matrix is set as
  \begin{equation}\label{Frf_P}
    {\bf{F}}_{\text{RF,P}} = \left(
                             \begin{array}{c}
                               \sqrt{\frac{K}{N_{\text{BS}}}} {\bf{U}}_{K \times K} \\
                               {\bf{1}}_{(N_{\text{BS}}-K) \times K}\\
                             \end{array}
                           \right)
  \end{equation}
where ${\bf{U}}$ is the $K \times K$ DFT matrix and the $K+1$ to $N_{\text{BS}}$ rows of ${\bf{F}}_{\text{RF,P}}$ are one padded. Of course, the elements of ${\bf{F}}_{\text{RF,P}}$ are distorted by the imperfect phase shifters but for simplicity of notation, we just drop the subscript E below.

   The received signal at the $k$-th user is expressed as
  \begin{equation*}\label{*}
  \begin{split}
    {\bf{y}}_{k} & = \sqrt{\rho}{\bf{w}}_{k}^*{\bf{H}}_{k} {\bf{F}}_{\text{RF,P}} + {\bf{w}}_{k}^* {\bf{n}}_{k} \\
    & =\sqrt{\frac{\rho v N_{\text{BS}} N_{\text{UE}}}{v+1}}{\bf{w}}_{k}^* {\bf{a}}_{{\text{UE}}} (\phi^{{\text{UE,L}}}_{k}) {\bf{a}}_{{\text{BS}}}^* (\phi^{{\text{BS,L}}}_{k}) {\bf{F}}_{\text{RF,P}} \\
   & + \sqrt{\frac{\rho N_{\text{BS}} N_{\text{UE}}}{L(v+1)}} {\bf{w}}_{k}^* \sum_{l=1}^{L} \gamma_{k,l} {\bf{a}}_{{\text{UE}}} (\phi^{{\text{UE,S}}}_{k,l}) {\bf{a}}_{{\text{BS}}}^* (\phi^{{\text{BS,S}}}_{k,l})  {\bf{F}}_{\text{RF,P}}  \\
     & \qquad + {\bf{w}}_{k}^* {\bf{n}}_{k} \\
          & = C^{{\text{L}}} {\bf{a}}_{{\text{BS}}}^* (\phi^{{\text{BS,L}}}_{k}){\bf{F}}_{\text{RF,P}} + \sum^{L}_{l=1} C^{{\text{S}}}_l  {\bf{a}}_{{\text{BS}}}^* (\phi^{{\text{BS,S}}}_{k,l}) {\bf{F}}_{\text{RF,P}}  \\
             & \qquad + {\bf{w}}_{k}^* {\bf{n}}_{k}\\
  & =  C^{{\text{L}}} \left[\sum_{n=1}^{K}e^{j(n-1)(0 - \phi^{{\text{BS,L}}}_{k})}  \sum_{n=1}^{K}e^{j(n-1)(\frac{2 \pi}{K} - \phi^{{\text{BS,L}}}_{k})}  \right. \\
  \end{split}
  \end{equation*}

  \begin{equation}\label{Received_at_th_user_channel_estimation}
     \begin{split}
          & \left. \qquad \cdots \sum_{n=1}^{K}e^{j(n-1)(\frac{2 \pi (K-1)}{K} - \phi^{{\text{BS,L}}}_{k})} \right]  \qquad \qquad \qquad\\
          & \qquad  + {\bar{\bf{n}}}_k  \\
     \end{split}
  \end{equation}
where $C^{{\text{L}}}=\sqrt{\frac{\rho v N_{\text{BS}} N_{\text{UE}} }{v+1}}{\bf{w}}_{k}^* {\bf{a}}_{{\text{UE}}} (\phi^{{\text{UE,L}}}_{k})$ and $C^{{\text{S}}}_l = \sqrt{\frac{\rho N_{\text{BS}} N_{\text{UE}} }{L(v+1)}} \gamma_{k,l}  {\bf{w}}_{k}^* {\bf{a}}_{{\text{UE}}} (\phi^{{\text{UE,S}}}_{k,l})$ are constants and not involved in the AOD estimation. ${\bar{\bf{n}}}_k$ is the equivalent noise including the contribution of training signals going through scattering paths. Each user uses only one omni-directional antenna to receive the $K$ pilot signals and the analog combining vector can be written as ${\bf{w}}_{k} = \frac{1}{\sqrt{N_{\text{UE}}}}[1,0,\cdots,0]^T$. ${\bf{y}}_k$ is the noisy complex-value DFT coefficients and the DFT size is $K$. Our goal is to estimate the downlink AOD of the LOS path $\phi^{{\text{BS,L}}}_{k}$. Quite a few algorithms of DFT interpolation have been published {\cite{Quinn,Jacobsen,Ligges}} and there is a comparison of them in {\cite{compare_DFT_interpolation}}. Finally we choose Jacobsen's algorithm for its better performance in low SNR regime.

 In Jacobsen's algorithm, we first find the maximum in the absolute value of the elements of ${\bf{y}}_{k}$ and its index is set as $k_{\text{max}}$. The indices of its two neighbors are $k_{\text{max}}-1$ and $k_{\text{max}}+1$, respectively. Note that $y_{k,k_{\text{max}}}$ can be at one end of ${\bf{y}}_{k}$ and in these cases one of its neighbors is the other end of ${\bf{y}}_{k}$. Then the AOD $\phi^{{\text{BS,L}}}_{k}$ can be estimated by the DFT interpolation of them three complex-value DFT coefficients

  \begin{equation}\label{Jacobsen}
    {\hat{\phi}}^{{\text{BS,L}}}_{k} = \frac{2 \pi k_{\text{max}} }{K} + \frac{2 \pi}{K} {\text{Re}} \left( \frac{y_{k,k_{\text{max}}-1} - y_{k,k_{\text{max}}+1}}{2y_{k,k_{\text{max}}} - y_{k,k_{\text{max}}+1} -y_{k,k_{\text{max}}-1}}\right)
  \end{equation}
  where $y_{k,i}$ is the $i$-th element of ${\bf{y}}_{k}$.

 Then it comes to the second stage of the user scheduling and analog beamformer design. The users feed back the estimated AOD ${\hat{\phi}}^{{\text{BS,L}}}_{k}$ to the BS. And if there are some AODs between which the difference is smaller than the beam width $2 \pi/N_{\text{BS}}$, they are scheduled as silent ones except one and the analog beamforming vectors corresponding to the silent users are zero vectors. The analog beamforming vector corresponding to the $k_I$-th non-silent user ${\bf{f}}_{\text{RF},{k_I}}$ can be set as $[1,e^{j{\mathcal{Q}} ( {\hat{\phi}}^{{\text{BS,L}}}_{k_I} )},e^{j{\mathcal{Q}} ( 2{\hat{\phi}}^{{\text{BS,L}}}_{k_I} )},\cdots,e^{j{\mathcal{Q}} ((N_{\text{BS}}-1) {\hat{\phi}}^{{\text{BS,L}}}_{k_I} )}]^T$.

Considering the multiple-antenna user equipment, it needs an extra stage to design the analog beamformer at the users. Restricted by the volume and power consumption, the amount of the antennas at the users is far less than the BS thus we can search for the best analog beamforming vector in the beamsteering codebook exhaustedly as {\cite{ZF_important}} and it won't take much overhead. Of course, unavoidably, the analog beamforming vectors will be distorted due to the phase-shifting error and gain error of phase shifters and we leave it to the digital beamformer.

At the third stage, we estimate the equivalent channel in the downlink. The BS sends orthogonal $K_I$-length pilot sequences to the users with ${\bf{F}}_{\text{RF,E}}$, which is derived at the second stage but distorted by the imperfect phase shifters, as the analog precoding matrix and ${\bf{w}}_{\text{E},k_I}$ as the analog combining vector. For simplicity, the pilot matrix is set as $\sqrt{\rho}{\bf{I}}_{K_I}$. At the $k_I$-th user, the received sequence is
\begin{equation}\label{equivalent_est_receive}
  {\bf{y}}_{k_I} = \sqrt{\rho}{\bf{w}}_{\text{E},k_I}^* {\bf{H}}_{k_I} {\bf{F}}_{\text{RF,E}} +  {\bf{w}}_{\text{E},k_I}^* {\bf{z}}_{k_I}
\end{equation}
 and we use ${\hat{{\bf{h}}}}_{\text{eq},k_I} = \frac{1}{\sqrt{\rho}}{\bf{y}}_{k_I}$ as the estimation of ${\bf{h}}_{\text{eq},k_I}$, which incurs the noise, inevitably.

In {\cite{ZF_important}}, they assume that the channel is reciprocal and estimate the equivalent channel in the uplink. But as we mentioned above, channel reciprocity can not be guaranteed and the synchronization among the users may cause extra overhead, thus we estimate the equivalent channel in the downlink and then feed back it. However, in some of other works, once they get the knowledge of propagation channel and design the hybrid beamformer according to it, they just use ({\ref{Heq}}) as the equivalent channel {\cite{No_Heq_estimation}} and they are very sensitive to the phase-shifting error and gain error. Although the estimation of equivalent channel may induce additional noise, it is beneficial to weaken the harm of the imperfect phase shifters and therefore enhance the performance as shown in the simulations. So there may be a tradeoff intuitively that in the low SNR regime where the harm of the noise is more serious than the imperfect phase shifters, the estimation of equivalent channel is more harm than good and can be omitted.

At the last stage, the estimated equivalent channel ${\hat{{\bf{h}}}}_{\text{eq},k_I}$ is fed back to the BS and the digital beamforming matrix can be set as ${\bf{F}}_{\text{BB}} = {\hat{\bf{H}}}_{\text{eq}}^* ({\hat{\bf{H}}}_{\text{eq}}{\hat{\bf{H}}}_{\text{eq}}^*)^{-1}$. In some systems with limited feedback requirement, the equivalent channel can be chosen among a codebook and only the index should be fed back{\cite{Limited_feedback}}.

\begin{algorithm}[htp]
\SetKwInOut{Input}{Require}
\Input{The pilot matrix from the BS to the users for the AOD estimation ${\bf{F}}_{\text{RF,P}}$, the orthogonal pilot sequences from the users to the BS for the equivalent channel estimation.}

{{{\bf{Stage 1. AOD estimation at the users}}}} \\

{\nl{The BS sends pilot matrix to the users with ${\bf{F}}_{\text{RF,P}}$ as the analog beamforming matrix. And the signal vector
    received at the $k$-th user is ${\bf{y}}_k$}}\;
{\nl{The $k$-th user estimates the AOD ${\hat{\phi}}^{{\text{BS,L}}}_{k}$ with ${\bf{y}}_k$ using Jacobsen's
    DFT interpolation algorithm which is shown in ({\ref{Jacobsen}})}}\;

{{{\bf{Stage 2 User scheduling and ABF design at the BS}}}} \\
{\nl{The users feed back the estimated AODs to the BS and the users within the AOD range of $2 \pi /N_{\text{BS}}$ are scheduled to be silent except one}}\;
{\nl{The phase of each element of the analog beamforming vector ${\bf{f}}_{\text{RF},k_I}$
is set as the quantization of that of the estimated antenna array response vector ${\bf{a}}_{\text{BS}} ({\hat{\phi}}^{{\text{BS,L}}}_{k_I} )$}} \;

{{{\bf{Stage 3 Equivalent channel estimation at the users}}}} \\
{\nl{The BS sends the orthogonal pilots to the users and the users estimate the equivalent channel
${\hat{{\bf{h}}}}_{\text{eq},k_I}$}}  \;

{\bf{Stage 4 DBF design at the BS}}  \\
{\nl{The users feed back ${\hat{{\bf{h}}}}_{\text{eq},k_I}$ to the BS and the digital beamforming matrix is set as the ZF precoder of the estimated equivalent channel
${\bf{F}}_{\text{BB}} = {\hat{{\bf{H}}}}_{\text{eq}}^*({\hat{{\bf{H}}}}_{\text{eq}}{\hat{{\bf{H}}}}_{\text{eq}}^*)^{-1}$}}  \;

 \caption{Channel estimation and HBF design algorithm}
\end{algorithm}

Next, we will compare our proposed algorithm with others in the following aspects. And we will show that our proposed algorithm outperforms some of the other algorithms, especially in {\cite{ZF_important}} conditionally with less training overhead and hardware cost. We also propose a way to further enhance the performance of our algorithm.

{
\subsection{Resolution of phase shifters}
Referring to {\cite{ZF_important}}, to guarantee the analog beamforming performance, the analog beamforming codebook at the BS should contain $J \approx \frac{2N_{\text{BS}}}{1.782}$ candidate vectors at least. If the beam steering vectors are used as candidate vectors, as {\cite{OMP,ZF_important}}, the digital phase shifters should have the resolution of ${\text{log}}_2 N_{\text{BS}}$-bits. In the massive MIMO systems, the BS is equipped with hundreds of antennas and the heavy use of the high resolution phase shifters may result in huge power consumption and hardware cost. Thus it is expected to use low complexity and power-saving phase shifters with few quantization bits. Alternatively, as proven in {\cite{SVD}}, if we get the optimal full digital beamformer ${\bf{F}}_{\text{opt}}$ at first, then extract the phase of the elements, and finally quantize them to obtain the phase of phase shifters, only $2$ or $3$-bits phase shifters are good enough no matter how many the antennas are. However, in {\cite{SVD}}, it should do SVD to the channel matrix to obtain ${\bf{F}}_{\text{opt}}$ and the computational complexity of SVD is too high for the hardware DSP. What's more, the whole knowledge of channel matrix consisting of both the path angles and the path coefficients is hard to obtain. In our algorithm, we only need to estimate the AOD of the LOS path and ${\bf{F}}_{\text{RF,opt}}$ can be obtained thus we can use low resolution phase shifters to approximate the phase of the elements of ${\bf{F}}_{\text{RF,opt}}$.

\subsection{The number of users supported}
 Referring to {\cite{compare_DFT_interpolation}}, the Jacobsen's estimator asymptotically holds
\begin{equation*}\label{DFT_size}
 N^{3/2} ({\hat{\theta}} - \theta) \sim {\mathcal{N}}(0,\frac{48 \pi f(\theta)}{A^2})
\end{equation*}
where $N$ is the DFT size, $f(\theta)$ is the spectral density of the noise. This reveals that with larger DFT size, the accuracy of the estimation can be improved qualitatively fixing the power of noise. And in our algorithm, it implies that with more users, the accuracy of the estimation is enhanced. To achieve the same performance with {\cite{ZF_important}} when the users are few, we can use several training cycles to enlarge the DFT size. The whole analog beamforming matrix in the AOD estimation stage containing $P$ training cycles can be expressed as

  \begin{equation*}\label{Frf_P_double}
    {\bf{F}}_{\text{RF,P}} = \left(
                             \begin{array}{c}
                               \sqrt{\frac{PK}{N_{\text{BS}}}} {\bf{U}}_{PK \times PK} \\
                               {\bf{1}}_{(N_{\text{BS}}-PK) \times PK}\\
                             \end{array}
                           \right)
  \end{equation*}

 And at the $p$-th cycle, we use the $(p-1)K+1$ to $pK$ columns of ${\bf{F}}_{\text{RF,P}}$ as the analog beamforming matrix, $p \in \{1,2,\cdots,P\}$. At the user side, the AOD estimation will be operated among the $PK$ received symbols thus the DFT size is $PK$ and we can make a tradeoff between the performance and the training overhead.

\subsection{The number of channel paths and Rice K-factor}
Considering Rice channel in mmWave systems, as we mentioned above, the analog beamforming vector is designed with the intention to steer the transmitting beam at the LOS path. So at the AOD estimation stage, the training signal going through the scattering paths will be regarded as noise at the user side. With more channel paths, and smaller Rician K-factor, the power of the noise is relatively larger thus the estimation will be more inaccurate. The algorithm in {\cite{ZF_important}} is robust in different propagation environment with different number of channel paths and Rician K-factor. To enhance the accuracy of estimation in this case, we can also use several training cycles to enlarge the DFT size thus to resist noise.

\subsection{Computational complexity and training overhead}
As mentioned above, the computational complexity of exhaustedly search in the codebook will increase linearly with $N_{\text{BS}}$ at the BS in the {\cite{ZF_important}}, as well as the training overhead. However, in our proposed algorithm, there is no need of the search process and the complexity of Jacobsen's estimation method increases linearly with the number of the users, as well as the training overhead. What's more, it must take several guard time for the stability of the phase shifters after changing their phases and the training time slots will be much longer using the algorithm in {\cite{ZF_important}}. In our algorithm, in the AOD estimation stage, the phase of the phase shifters need not be changed, which will save the guard time as well as the power. In the cases where we should take several cycles to get better performance, the training overhead is multiplying the number of cycles by the number of users, which is also far less than the number of antennas. Of course the guard time is also required.

}

\begin{rem}
In the AOD estimation stage, if the analog beamforming matrix is
  \begin{equation*}\label{Frf_P_double}
    {\bf{F}}_{\text{RF,P}} = \left(
                             \begin{array}{c}
                               \sqrt{\frac{K}{N_{\text{BS}}}} {\bf{U}}_{K \times K} \\
                               {\bf{0}}_{(N_{\text{BS}}-K) \times K}\\
                             \end{array}
                           \right)
  \end{equation*}
the accuracy of AOD estimation will be greatly improved with the existence of phase-shifting error and gain error. However the elements `0' can not be realized by phase shifters. Considering the hardware structure in {\cite{SWITCH_AND_PS}} where each phase shifter is followed by a switch, we can realize the elements `0' by turning off the switch. This may cause extra hardware cost but it can use less training cycles and less phase shifters to achieve the same performance, which may consume less power.

\end{rem}
%
%
%

 \section{Simulations}
  In this section we will show our simulation results about the degradation of the performance with phase-shifting error and gain error in multiuser scenario. And the performance of our proposed algorithm will also be shown later, as well as the comparison with others.

  \subsection{Performance loss due to phase-shifting error and gain error}
  In the downlink multiuser scenario, we set the number of the antennas at the BS as $N_{\text{BS}} = 128$  and each user is equipped with $N_{\text{UE}} = 4$ antennas. Both the BS and the users have ULA antennas. The channel between the BS and each user is assumed to be Rice model with $4$ paths totally and the Rician K-factor is $30$. The channel knowledge is assumed to be perfect both at the BS and the user side. The number of the total users is $K $. Each user is served by a single RF chain thus the number of RF chains at the BS equals $K$.  As to the hybrid beamforming process, the analog beamforming matrix is set based on the SVD of channels by extracting the phase of the elements of the right and left singular vector corresponding to largest singular value, and then quantize them with $3$-bits phase shifters. The digital precoder at the BS is ZF precoder based on the equivalent channel in ({\ref{Heq}}). The transmit signal-to-noise ratio is defined as ${\text{SNR}} = \rho / \sigma^2_z$.

\begin{figure}[ht]
  \centering
  \includegraphics[width=0.7\textwidth]{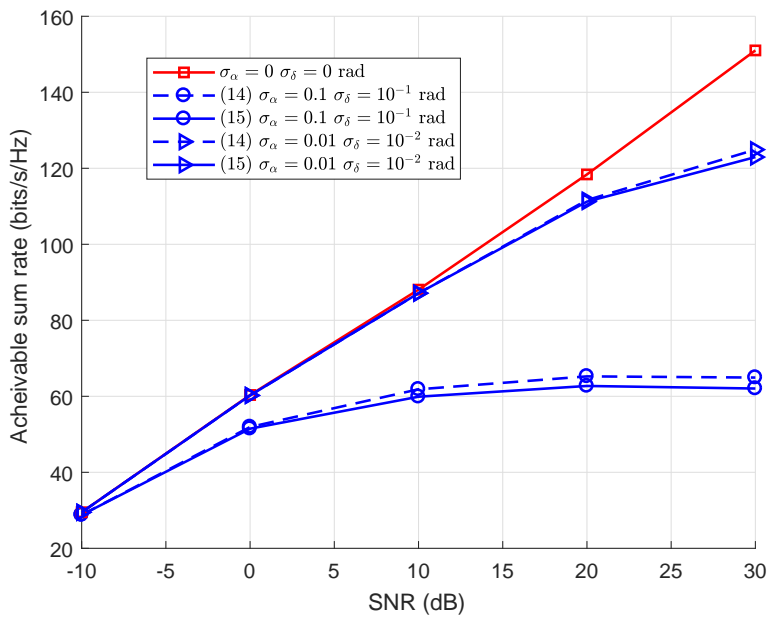}\\
  \caption{Achievable sum rate with imperfect phase shifter according to (14) and (15). }\label{aMultiuser_13_14}
 \end{figure}

  As we mentioned before, the upper bound of achievable sum rate with imperfect phase shifters can be derived according to ({\ref{rate error_mean}}), which is an approximation, exactly lower bound, of the practical achievable sum rate shown in ({\ref{rate_error}}). In Figure {\ref{aMultiuser_13_14}}, we can see that there is only a very small gap between ({\ref{rate error_mean}}) and ({\ref{rate_error}}). Therefore in the following figures, we use ({\ref{rate error_mean}}) as the simulation result of achievable sum rate with imperfect phase shifters.

\begin{figure}[ht]
  \centering
  \includegraphics[width=0.7\textwidth]{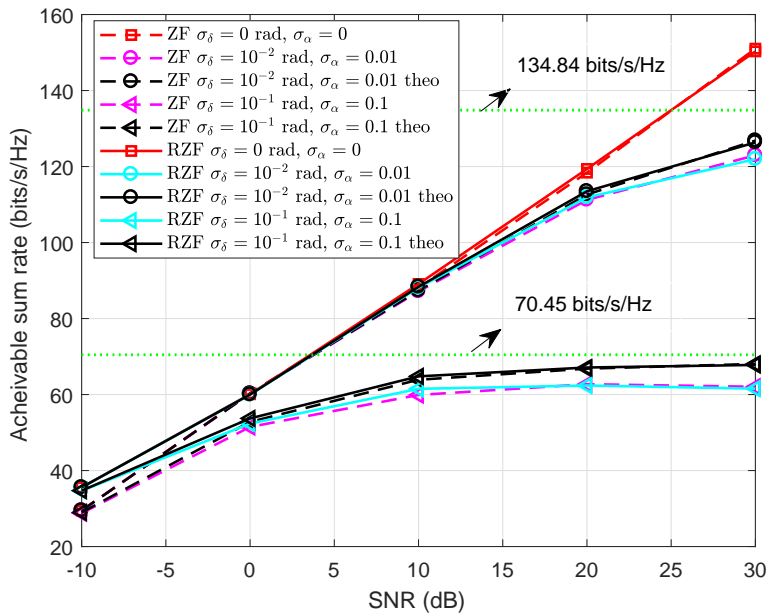}\\
  \caption{Achievable sum rate versus SNR different rms values of phase-shifting error and gain error.}\label{aMultiuser_ONLY_PHASE_ERROR}
\end{figure}

  Then we investigate the effect of phase-shifting error and gain error jointly in Figure {\ref{aMultiuser_ONLY_PHASE_ERROR}}. The number of the users is $10$. We can see that when $\sigma_{\alpha} = 0.1$ and $\sigma_{\delta} = 10^{-1}$rad, the loss ratio of the achievable sum rate is nearly $50\%$ when ${\text{SNR}}=25$ dB. With RZF digital precoder, the achievable sum rate can be slightly improved in the low SNR regime while in the high SNR regime, there is no difference between the performance of ZF and RZF digital precoder and they are all limited to the performance ceiling shown by the green dotted lines derived by formula ({\ref{Upperbound_128_10}}). The theoretical upper bound shown by the black lines according to formula ({\ref{rate error_mean}}) and ({\ref{lower bound final}}) is very tight and it is suitable for the RZF digital precoder, too.

\begin{figure}[ht]
  \centering
  \includegraphics[width=0.7\textwidth]{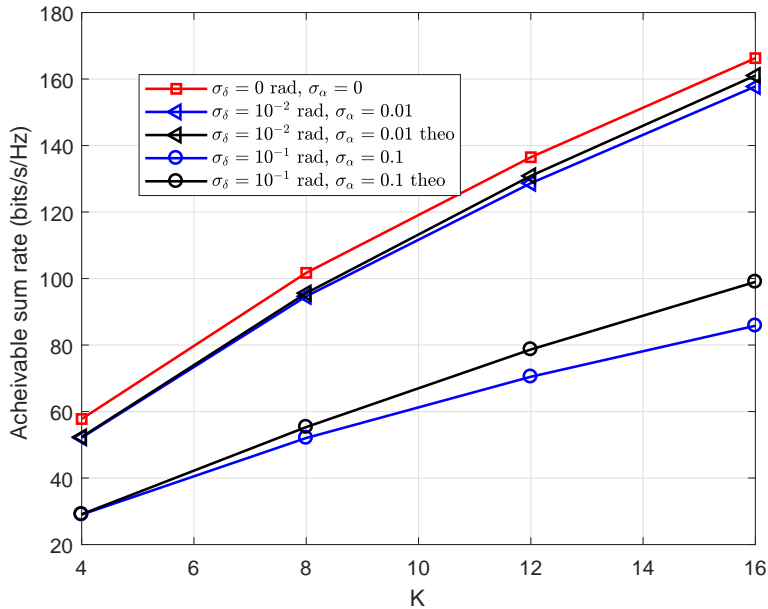}\\
 \caption{Achievable sum rate versus $K$ with different rms value of phase-shifting error and gain error. }\label{aMultiuser_ONLY_GAIN_ERROR}
\end{figure}

In Figure {\ref{aMultiuser_ONLY_GAIN_ERROR}}, we fix the SNR as $20$ dB and change the number of users from $4$ to $16$. With larger amount of users, the orthogonality among users can not be guaranteed and the inter-user interference could be more serious. Thus the theoretical upper bound shown by black lines is tighter when the user is less for the orthogonality can be satisfied. Applying our user scheduling method mentioned in section \uppercase\expandafter{\romannumeral4}, some of the users may be silent and the achievable sum rate does not increase linearly with the number of users. And the gain of achievable sum rate with more users further drops when there exist the phase-shifting error and gain error.

\subsection{Performance of our proposed channel estimation and hybrid beamforming algorithm}
In this section we will compare the performance of several existing algorithms with our proposed one with respect of different parameters. In the following three figures, the curves in blue and green both use the SVD based HBF design algorithm with the assumption of full knowledge of CSI. The analog beamforming vectors are obtained by doing SVD to the channel as we mentioned in section \uppercase\expandafter{\romannumeral3}, and the ZF digital precoder is based on the equivalent channel in ({\ref{Heq}}). This algorithm is similar to {\cite{No_Heq_estimation}} and the only difference is that they proposed a low complexity algorithm to design ${\bf{F}}_{\text{RF}}$ therein. However, the curves in blue show the achievable sum rate with perfect phase shifters, which can be regarded as the benchmark while the curves in green represent the performance with imperfect phase shifters. The curves in red are corresponding to the HBF design algorithm in {\cite{ZF_important}}, where the analog beamforming vectors are selected from the beamsteering codebook and the digital precoding matrix is ZF precoder based on the estimated equivalent channel. The curves in black show the performance of our proposed channel AOD estimation and hybrid beamforming design algorithm. The number of antennas at the users is $4$ and the resolution of the phase shifters is $\frac{2 \pi}{N_{\text{BS}}}$ at the BS and $\frac{2 \pi}{N_{\text{UE}}}$ at the user side.

\begin{figure}[ht]
  \centering
  \includegraphics[width=0.7\textwidth]{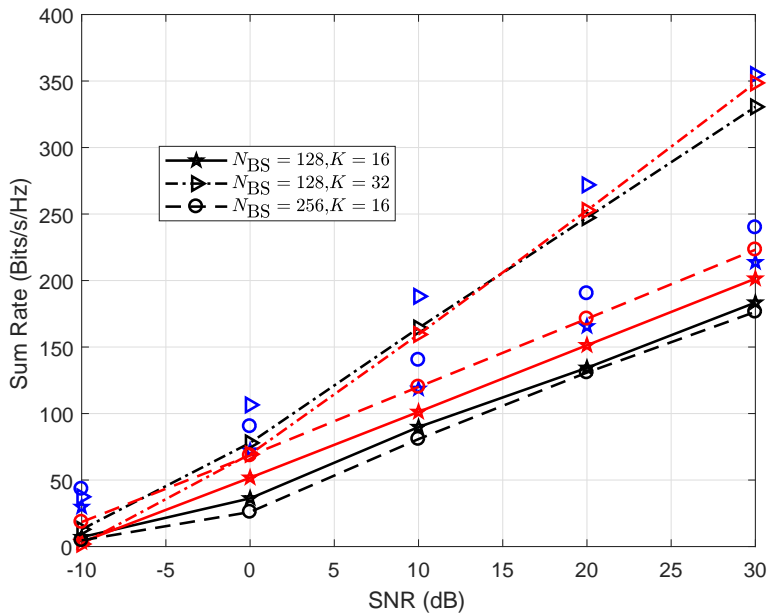}\\
  \caption{Achievable sum rate versus SNR with different number of users and the antennas at the BS.  Rician K-factor $v = 30$, $2$ channel paths, $\sigma_{\alpha} = 0.01$ and $\sigma_{\delta} = 10^{-2}$ rad. }\label{NT_BS_K}
\end{figure}

In Figure {\ref{NT_BS_K}}, it is interesting to see that with the same number of users $K=16$, the performance of our proposed algorithm with $N_{\text{BS}}=256$ is a little worse than that with $N_{\text{BS}}=128$ and the performance gap between the algorithm in {\cite{ZF_important}} and ours is also larger when $N_{\text{BS}}=256$. When $N_{\text{BS}} = 128$, $K = 32$, our proposed algorithm outperforms the algorithm in {\cite{ZF_important}} in the low regime. The reason is that with relatively larger DFT size, the accuracy of AOD estimation can be improved. While in the high regime, the algorithm in {\cite{ZF_important}} turns the tide. It can be explained that in {\cite{ZF_important}} the probability of the number of non-silent users $p(K_I)$ is different from ours. In {\cite{ZF_important}}, if more than one users share the same analog precoding vector, which is selected among the total $N_{\text{BS}}$ candidate vectors, they are scheduled as silent users except one. By Monte-Carlo simulation, the average of the non-silent users is $28.42$ when $N_{\text{BS}} = 128$, $K = 32$ and $15.10$ when $N_{\text{BS}} = 128$, $K = 16$. However, in our proposed algorithm, the analog precoding vectors are designed based on the AOD of the strongest path and the AODs are uniformly distributed in $[0,2 \pi)$. By Monte-Carlo simulation, the average of the non-silent users is $25.09$ when $N_{\text{BS}} = 128$, $K = 32$ and $14.22$ when $N_{\text{BS}} = 128$, $K = 16$. Thus we can draw the conclusion that the average number of non-silent users using the algorithm in {\cite{ZF_important}} is larger than our proposed algorithm and the SVD based HBF algorithm, which will result in higher achievable sum rate.

\begin{figure}[ht]
  \centering
  \includegraphics[width=0.7\textwidth]{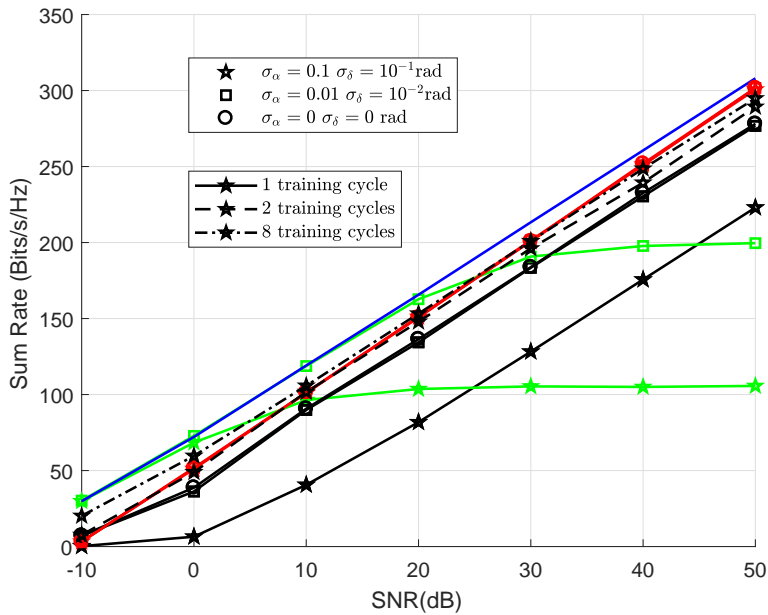}\\
  \caption{Achievable sum rate versus SNR with different phase-shifting error and gain error. The number of the antennas at the BS is $N_{\text{BS}} = 128$, the number of the users is $K=16$, Rician K-factor $v = 30$, $2$ channel paths.}\label{GAIN_PHASE_ERROR}
\end{figure}

In Figure {\ref{GAIN_PHASE_ERROR}}, we can see that the performance in {\cite{No_Heq_estimation}} is very sensitive to the gain error and phase-shifting error. In the low SNR regime, it performs best while in the high SNR regime, the performance will be limited to the ceiling, which indicates the necessity of the estimation of the equivalent channel. Both of the other two algorithms can remove the performance ceiling. The red curves corresponding to different level of phase-shifting error and gain error all coincide, which indicates that the algorithm in {\cite{ZF_important}} is the most robust with phase-shifting error and gain error. When $\sigma_{\alpha} = 0.1$ and $\sigma_{\delta} = 10^{-1}$ rad, the accuracy of AOD estimation drops seriously and the performance of our algorithm fall behind greatly while when $\sigma_{\alpha} = 0.01$ and $\sigma_{\delta} = 10^{-2}$ rad, the influence of phase-shifting error and gain error can be neglected. However there is a performance gap between the red and black curves even with perfect phase shifters because the accuracy of AOD estimation is limited due to the training signals going through the scattering paths, which can be regarded as part of the equivalent noise. Then we focus on the condition that $\sigma_{\alpha} = 0.1$ and $\sigma_{\delta} = 10^{-1}$ rad and use multiple training cycles, as shown by the dash line with $2$ training cycles and the dot dash line with $8$ training cycles, to enhance the performance. We can see that with $2$ training cycles, the performance of our proposed algorithm is almost the same as the algorithm in {\cite{ZF_important}} when SNR is less than $30$ dB. And with $8$ training cycles, which takes the same training overhead as the algorithm in {\cite{ZF_important}}, our proposed algorithm performs better.

\begin{figure}[ht]
  \centering
  \includegraphics[width=0.7\textwidth]{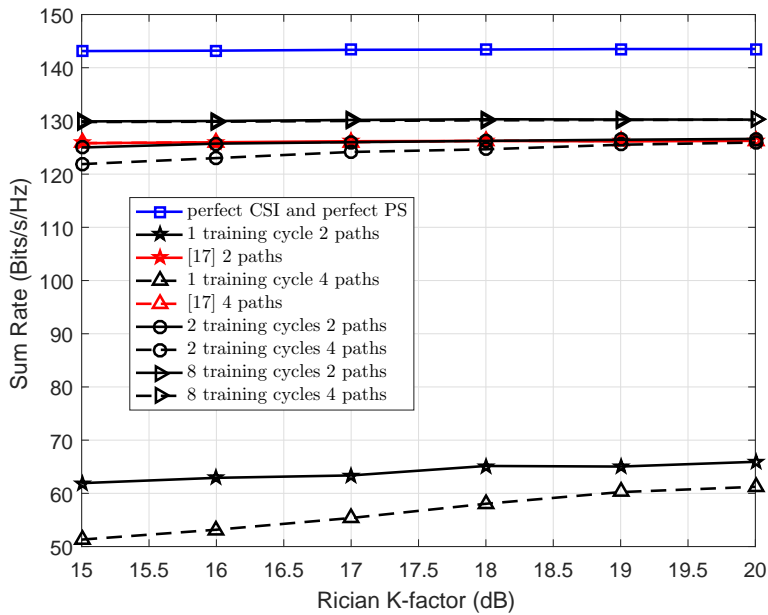}\\
  \caption{Achievable sum rate versus Rician K-factor with different number of the channel paths. The number of the antennas at the BS is $N_{\text{BS}} = 128$, the number of the users is $K=16$, SNR is $15$ dB, $\sigma_{\alpha} = 0.1$ and $\sigma_{\delta} = 10^{-1}$ rad. }\label{RF}
\end{figure}

Then we fix SNR as $15$ dB and look into the influence of the Rician K-factor upon the achievable sum rate. As shown in Figure {\ref{RF}}, the algorithm in {\cite{ZF_important}} is robust in different propagation environment. And when $\sigma_{\alpha} = 0.1$ and $\sigma_{\delta} = 10^{-1}$ rad, our proposed algorithm using only one training cycle suffers a serious performance loss. With more channel paths and smaller Rician K-factor the performance further drops. Considering the training overhead, the algorithm in {\cite{ZF_important}} should search the codebook exhaustedly and the training time slots in need are $N_{\text{BS}}+N_{\text{UE}}$ to form the analog beamforming matrix while our proposed algorithm demands only $K+N_{\text{UE}}$ time slots. Then we make a tradeoff between the performance and the overhead. With two training cycles, the performance can be greatly improved and there is only a small performance gap between the algorithm in {\cite{ZF_important}} and ours. With eight training cycles, the performance is further enhanced and better than the algorithm in {\cite{ZF_important}}.

To sum up, thanks to the few propagation paths and the strong LOS path in multiuser mmWave communication systems, the performance of our proposed algorithm is not worse than the algorithm in {\cite{ZF_important}}, sometimes even better. And the performance ceiling due to the imperfect phase shifters can also be removed. Meanwhile the training overhead can be greatly saved.
\section{Conclusion}
 In this paper, we looked into the loss of achievable sum rate in multiuser scenario caused by phase-shifting error and gain error of phase shifters. We derived the upper bound of the achievable sum rate and the expression is validated via simulations. The theoretical and simulation results indicate that the degradation of the performance is serious with the existence of phase-shifting error and gain error and will be limited to the ceiling in the high SNR regime. Some compensation algorithms may be required to deal with phase-shifting error and gain error. Then we proposed a downlink channel estimation and hybrid beamforming algorithm, which can enjoy good performance with much lower training overhead. The tradeoff between the performance and training overhead is further made to adjust the algorithm to more communication conditions.

\section*{Appendix A\\Proof of the Theorem 1}

 {\flushleft{\emph{Proof:} The derivation of the lower bound is based on the assumption that the resolution of phase shifters is high enough and $N_{\text{BS}} \to \infty$. Firstly we do SVD to the channel between the BS and $k$th user, and the data stream is desired to be transmitted through the strongest channel path. We have }}
\begin{equation}\label{channel_k}
  {\bf{h}}_{{\text{eq}},k_I} =  {\bf{w}}^*_{k_I} {\bf{u}}_{k_I} \sigma_{k_I} {\bf{v}}^*_{k_I} {\bf{F}}_{\text{RF}}
\end{equation}
 where $\sigma_k$ is the largest singular value of ${\bf{H}}_{k_I}$ and ${\bf{u}}_{k_I}, {\bf{v}}^*_{k_I}$ is the left and right singular vector corresponding to $\sigma_k$. ${\bf{u}}_{k_I}, {\bf{v}}^*_{k_I}$ are the linear combinations of the antenna array response vectors of AOA and AOD {\cite{OMP}}. When the resolution of phase shifters is high enough, we have  ${\bf{w}}^*_{k_I} {\bf{u}}_{k_I} = 1$ and ${\bf{v}}^*_{k_I} {\bf{f}}^{\text{RF}}_{k_I} = 1$. When $N_{\text{BS}} \to \infty$, the right singular vector corresponding to the largest singular value of each user is orthogonal to each other ${\bf{v}}^*_{k_I} {\bf{v}}_{u} = 0$ thus ${\bf{v}}^*_{k_I} {\bf{f}}^{\text{RF}}_{u} = 0$.

 Based on these assumption, the equivalent channel is approximate to a diagonal matrix and can be expressed as ${\bf{H}}_{\text{eq}} = {\text{diag}}(\sigma_1,\sigma_2,\cdots,\sigma_{K_I})$.

 And the ZF precoder at the BS is given as ${\bf{F}}_{\text{BB}} = {\text{diag}}( 1/{\sigma_1}, 1/{\sigma_2},\cdots, 1/{\sigma_{K_I}})$.

 When there exist the phase-shifting error and gain error, we define $\epsilon^{{\text{UE}}}_{k_I} = {\bf{w}}^{*}_{{\text{E}},k_I} {\bf{u}}_{k_I} =1/{{N_{\text{UE}}}} \sum^{N_{\text{UE}}}_{i=1} \alpha_{k_I i} e^{j \delta_{k_I i}}$
  and
   $\epsilon^{{\text{BS}}}_{k_I} = {\bf{f}}^{{\text{RF}}*}_{{\text{E}},k_I} {\bf{v}}_{k_I} = 1/{{N_{\text{BS}}}} \sum^{N_{\text{BS}}}_{i=1} \alpha_{k_I i}e^{j \delta_{k_I i}}$
   and
   $\xi_{k_I u} = {\bf{f}}^{{\text{RF}}*}_{{\text{E}},k_I} {\bf{v}}_{u} =1/{{N_{\text{BS}}}} \sum^{N_{\text{BS}}}_{i=1} \alpha_{k_I i} e^{j\delta_{k_I i}} e^{j(\theta^{{\text{BS}}}_{ui} - \theta^{{\text{BS}}}_{k_I i})}$. With $\delta \sim {\mathcal{N}}(0,\sigma^2_\delta)$ and  $\alpha \sim {\mathcal{N}}(1,\sigma^2_\alpha)$, we have
\begin{small}
 \begin{subequations}\label{mean_results}
 \begin{align}
  E[|\epsilon^{{\text{UE}}}_{k_I} |^2] & = \frac{1}{N_{\text{UE}}^2} (E[\alpha_i^2]N_{\text{UE}} +(N_{\text{UE}}^2-N_{\text{UE}})E[\alpha_i \alpha_j]E[e^{j(\delta_i - \delta_j)}] ) \nonumber\\
   & = { (1+ \sigma_{\alpha}^2 -e^{-\sigma_{\delta}^2})}/{ N_{\text{UE}}} + e^{-\sigma_{\delta}^2} \\
 E[|\epsilon^{{\text{BS}}}_{k_I} |^2] & =  { (1+ \sigma_{\alpha}^2 -e^{-\sigma_{\delta}^2})}/{ N_{\text{BS}}} + e^{-\sigma_{\delta}^2} \\
 E[\epsilon^{{\text{UE}}}_{k_I} \epsilon^{{\text{BS}}}_{k_I} ] & = e^{-\sigma_{\delta}^2} \\
 E[|\xi_{k_I u}|^2] & = \frac{1}{N_{\text{BS}}^2}(E[\alpha_i^2]-E[\alpha_i \alpha_j]E[e^{j(\delta_{ki}-\delta_{kj})}]) \nonumber \\
 & \quad \cdot \sum^{N_{\text{BS}}}_{i=1} |e^{j(\theta^{{\text{BS}}}_{ui} - \theta^{{\text{BS}}}_{ki})}|^2 \nonumber \\
 & = {(1 + \sigma_{\alpha}^2 -e^{-\sigma_{\delta}^2})}/{N_{\text{BS}}}
 \end{align}
  \end{subequations}
  \end{small}
  Then we can express the difference between equivalent channel with perfect and imperfect phase shifters $\Delta {\bf{H}}_{\text{eq}} =$
\begin{equation}\label{F_BB_simple}
   \left[
                           \begin{array}{cccc}
                             (\epsilon^{\text{UE}}_{1} \epsilon^{\text{BS}}_{1}- 1){\sigma_1} & \small{\cdots} & \epsilon^{\text{UE}}_{1}\xi_{1 {K_I}}{\sigma_1} \\
                             \epsilon^{\text{UE}}_{2}\xi_{21}{\sigma_2} & \cdots & \epsilon^{\text{UE}}_{2}\xi_{2 {K_I} }{\sigma_2} \\
                             \vdots & \ddots & \vdots \\
                            \epsilon^{{\text{UE}}}_{K_I} \xi_{K_I 1}{\sigma_{K_I}} & \cdots & (\epsilon^{{\text{UE}}}_{K_I}  \epsilon^{{\text{BS}}}_{K_I} - 1){\sigma_{K_I}} \\
                           \end{array}
                         \right]
\end{equation}

 And then $ E_{\Delta {\bf{h}}_{{\textrm{eq}},k_I} }[\Delta {\bf{h}}_{{\textrm{eq}},k_I} {\bf{F}}_{\textrm{BB}}{\bf{F}}_{\textrm{BB}}^* \Delta {\bf{h}}_{{\textrm{eq}},k_I} ^* ] =$
\begin{equation}\label{hahahh}
\begin{aligned}
\begin{split}
& E \left[ |\epsilon^{{\text{UE}}}_{k_I}  \epsilon^{{\text{BS}}}_{k_I} -1|^2 + \sum^{K_I}_{u=1,u \neq k_I} \frac{|\epsilon^{{\text{UE}}}_{k_I} |^2|\xi_{k_I i}|^2\sigma_{k_I}^2}{\sigma_i^2} \right]\\
& =   E[|\epsilon^{{\text{UE}}}_{k_I} |^2] E[|\epsilon^{{\text{BS}}}_{k_I} |^2]-2 E[ {\emph{\text{Re}}} \{\epsilon^{{\text{UE}}}_{k_I}  \epsilon^{{\text{BS}}}_{k_I} \}]\\
& \quad +1 + (K_I-1) E[|\epsilon^{{\text{UE}}}_{k_I} |^2]  E[|\xi_{k_I u}|^2]
\end{split}
\end{aligned}
\end{equation}

Substitute ({\ref{mean_results}}) into ({\ref{hahahh}}), it results
\begin{equation}\label{lower bound}
\begin{aligned}
 E_{\Delta {\bf{h}}_{{\textrm{eq}},k_I} }[x_{k_I}] & = \left( \frac{K_I (1 + \sigma^2_{\alpha} -e^{-\sigma^2_{\delta}})}{ N_{\text{BS}}} + e^{-\sigma^2_{\delta}} \right) \\
  & \quad \times \left( \frac{1 + \sigma^2_{\alpha} -e^{-\sigma^2_{\delta}}}{ N_{\text{UE}}} +e^{-\sigma^2_{\delta}} \right)\\
& \quad -2e^{-\sigma^2_{\delta}}+ 1\\
\end{aligned}
\end{equation}

 When $N_{\text{BS}}$ is finite, the orthogonality of ${\bf{v}}_{k_I},k_I=1,2,\cdots,K_I$ may be not satisfied and the conclusion in ({\ref{lower bound}}) is the lower bound. We can use the approximation $1-e^{-\sigma^2_\delta} \approx \sigma^2_\delta$ to simplify the formula ({\ref{lower bound}}) as ({\ref{lower bound final}}) when the variance of phase-shifting error is smaller than $10^{-1}$ ${\text{rad}}^2$.     $\hfill\blacksquare$

\bibliographystyle{IEEEtran}
\bibliography{hhh}

\end{document}